\documentclass[prd,amsmath,nobibnotes,nofootinbib,superscriptaddress,preprintnumbers]{revtex4-1}

\usepackage{bm} \usepackage{graphicx}
\usepackage{epstopdf}
\graphicspath{{Figures/}}

\def\ba{\begin{eqnarray}}
\def\ea{\end{eqnarray}}
\def\be{\begin{equation}}
\def\ee{\end{equation}}
\def\({\left(}
\def\){\right)}
\def\[{\left[}
\def\]{\right]}
\def\<{\left<}
\def\>{\right>}

\begin{document}

\title{Escaping the crunch: gravitational effects in classical transitions}
\date{\today}

\author{Matthew C Johnson}
\email{mjohnson@theory.caltech.edu}
\affiliation{California Institute of Technology, Pasadena, CA 91125, USA} 
\author{I-Sheng Yang}
\email{isheng.yang@gmail.com}
\affiliation{ISCAP and Physics Department, Columbia University, New York, NY 10027, USA}

\begin{abstract}
During eternal inflation, a landscape of vacua can be populated by the nucleation of bubbles. These bubbles inevitably collide, and collisions sometimes displace the field into a new minimum in a process known as a classical transition. In this paper, we examine some new features of classical transitions that arise when gravitational effects are included. Using the junction condition formalism, we study the conditions for energy conservation in detail, and solve explicitly for the types of allowed classical transition geometries. We show that the repulsive nature of domain walls, and the de Sitter expansion associated with a positive energy minimum, can allow for classical transitions to vacua of higher energy than that of the colliding bubbles. Transitions can be made out of negative or zero energy (terminal) vacua to a de Sitter phase, re-starting eternal inflation, and populating new vacua. However, the classical transition cannot produce vacua with energy higher than the original parent vacuum, which agrees with previous results on the construction of pockets of false vacuum. We briefly comment on the possible implications of these results for various measure proposals in eternal inflation.
\end{abstract}

\preprint{CALT-68.2791}

\maketitle

\section{Introduction}

The ubiquity of vacua in theories beyond the standard model of particle physics has highlighted the importance of understanding how our particular low-energy vacuum was accessed in the early universe. The need for such an understanding is particularly urgent in theories with extra dimensions, such as string theory, where vast collections of low-energy vacua (features in a potential ``landscape") can result upon compactification. A central question is to determine how (or if) the various states in the landscape can be populated.

Typically, the vacua in a potential landscape will not be completely stable. Their decay occurs via the nucleation of bubbles containing some new phase, which subsequently expand into the parent vacuum. The process of bubble nucleation can in many cases be described by the Coleman de Luccia (CDL) instanton~\cite{Coleman:1977py,Coleman:1977th,Coleman:1980aw}. Starting in some positive energy minimum, the resulting inflationary expansion can end locally due to bubble nucleation events. However, when the rate of bubble formation is roughly less than one bubble per Hubble volume per Hubble time, the phase transition cannot complete, and inflation becomes eternal (for some reviews of eternal inflation, see~\cite{Guth:2007ng,Aguirre:2007gy}). Bubble nucleation during eternal inflation can populate many of the vacua in a landscape.

Including gravitational effects in the description of domain walls and vacuum bubbles leads to some dramatically new phenomena. One consequence is that domain walls become repulsive \cite{Ipser:1983db,Vilenkin:1984hy}. Another important, purely gravitational, effect is the existence of surviving bubbles containing false vacuum (a higher energy phase than the background). In the absence of gravity, the pressure gradient pointing into a false vacuum bubble will cause it to collapse to zero size. Including gravity, if the interior of the false vacuum bubble has positive energy, and is larger than the associated de Sitter horizon size, it cannot (by causality) collapse to zero size. 

Although the phase transition does not percolate, there will still be collisions between bubbles. In fact, since each bubble grows to reach infinite size, all bubbles will undergo an infinite number of collisions. Instead of dissipating the collision energy, the field configuration can remain compact and naturally form a domain wall enclosing a region of some other vacuum. Locally, the collision has kicked the field into a new basin of attraction. This represents a secondary mechanism for populating vacua during eternal inflation, and is referred to as a classical transition~\cite{Easther:2009ft}. Classical transitions have been studied in simulations without gravity~\cite{Easther:2009ft}, and their field dynamics in flat space further elaborated on in~\cite{GibHui10}.

These studies of classical transitions in flat space suggest that, in the thin-wall limit, the spacetime can be modeled as  vacuum regions connected across domain walls. Under this approximation, we can apply the Israel Junction condition formalism~\cite{Isr66} to determine the global structure of the spacetime, including gravitational effects. The junction condition formalism has been an important tool in determining the outcome of bubble collisions and the large-scale structure of the resulting spacetimes. Some representative examples in the literature include~\cite{HawMos82a,Wu:1984eda,Freivogel:2007fx,Aguirre:2007wm,Chang:2007eq}. 

In this paper, we explore the large-scale structure of classical transition geometries, determining the general criteria that must be met for a classical transition to occur, and outlining solutions where the effects of gravity are important. There are many parallels with the types of new solutions that arise for vacuum bubbles. One particularly interesting result is that classical transitions can give rise to a lasting region with a vacuum of higher energy than the interior of the colliding bubbles. This allows for the creation of inflating vacua from the collision of bubbles containing so-called terminal vacua, thereby providing a means of connecting otherwise isolated sectors of a landscape. We formulate constraints on the types of classical transitions that are allowed by energy conservation, and assess the types of vacua it is possible to populate using this mechanism.

The plan of the paper is as follows. In Sec.~\ref{sec:fielddynamics}, we briefly describe the field dynamics of classical transitions. We then describe how the Israel Junction condition formalism can be used to construct classical transition spacetimes in Sec.~\ref{sec:junctionconditions}. In Sec.~\ref{sec-dS}  we specialize to the collision of bubbles containing a positive or zero energy phase, classifying the possible classical transition geometries, and deriving a number of constraints from energy conservation. Some of these results are generalized and expanded upon in Appendix~\ref{app-oscillate} and~\ref{app-B1B2}. Colliding bubbles containing a negative energy phase allows for a wider variety of classical transition geometries, which we study separately in Sec.~\ref{sec-AdS}. Some technical results are collected in Appendix~\ref{app-turnaround}. Finally, we briefly discuss the implications of these results and conclude in Sec.~\ref{sec:conclusions}. 

\section{Classical transitions: field dynamics}\label{sec:fielddynamics}

Analytic studies of bubble collisions~\cite{BouFre06a,BouFre08a,Aguirre:2007wm,Chang:2007eq,Freivogel:2007fx,Wu:1984eda} depend heavily on the assumptions made about how colliding domain walls interact. Some motivated guesses for the possible outcome of a collision between two bubbles include the creation of shells of radiation or dust, and in the case of non-identical bubbles, a domain wall that separates the bubble interiors. Numerical simulations of bubble collisions in flat space~\cite{HawMos82a,Aguirre:2008wy} have confirmed these types of behavior, but also revealed a number of new possibilities. Simulations by Easther et al~\cite{Easther:2009ft} uncovered the existence of a phenomenon known as the ``classical transition,'' whose study will be the main focus of this paper. Progress in understanding the scalar field dynamics, presented in a recent paper~\cite{GibHui10}, will allow us to set up our analytic study. We quickly review the main results of this study here.

\begin{figure}[tb]
   \includegraphics[width=6 cm]{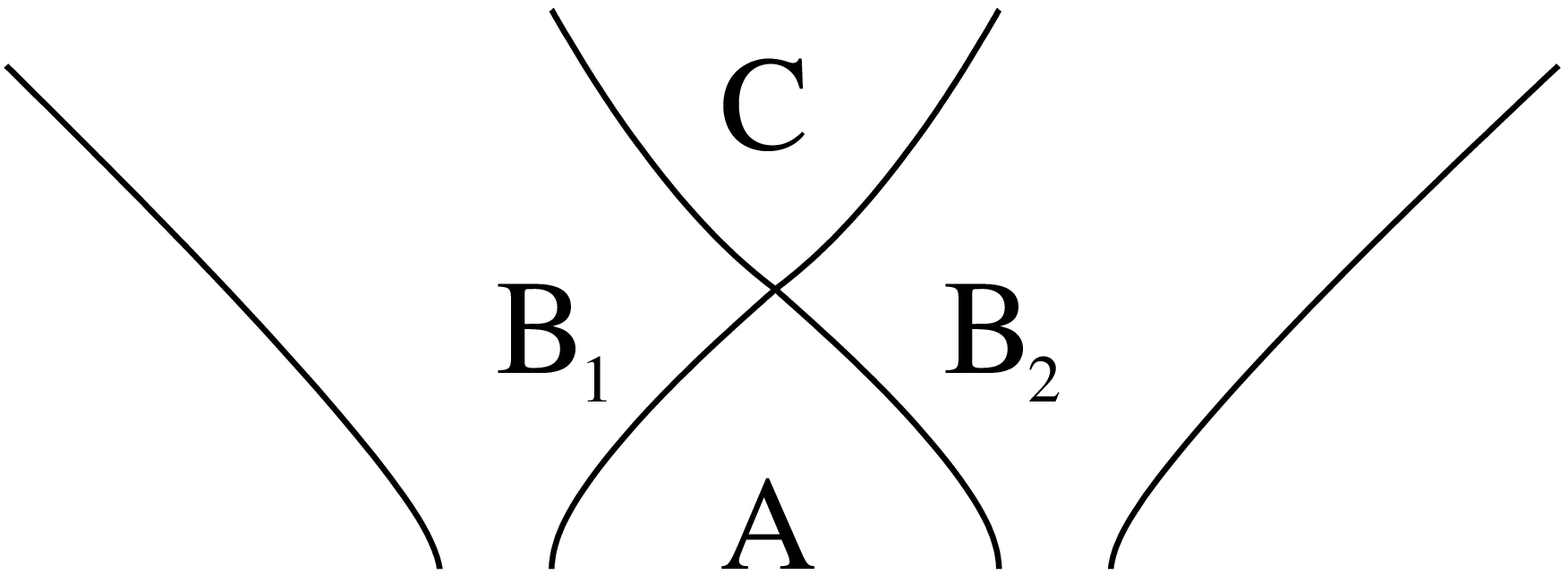}
   \includegraphics[width=6 cm]{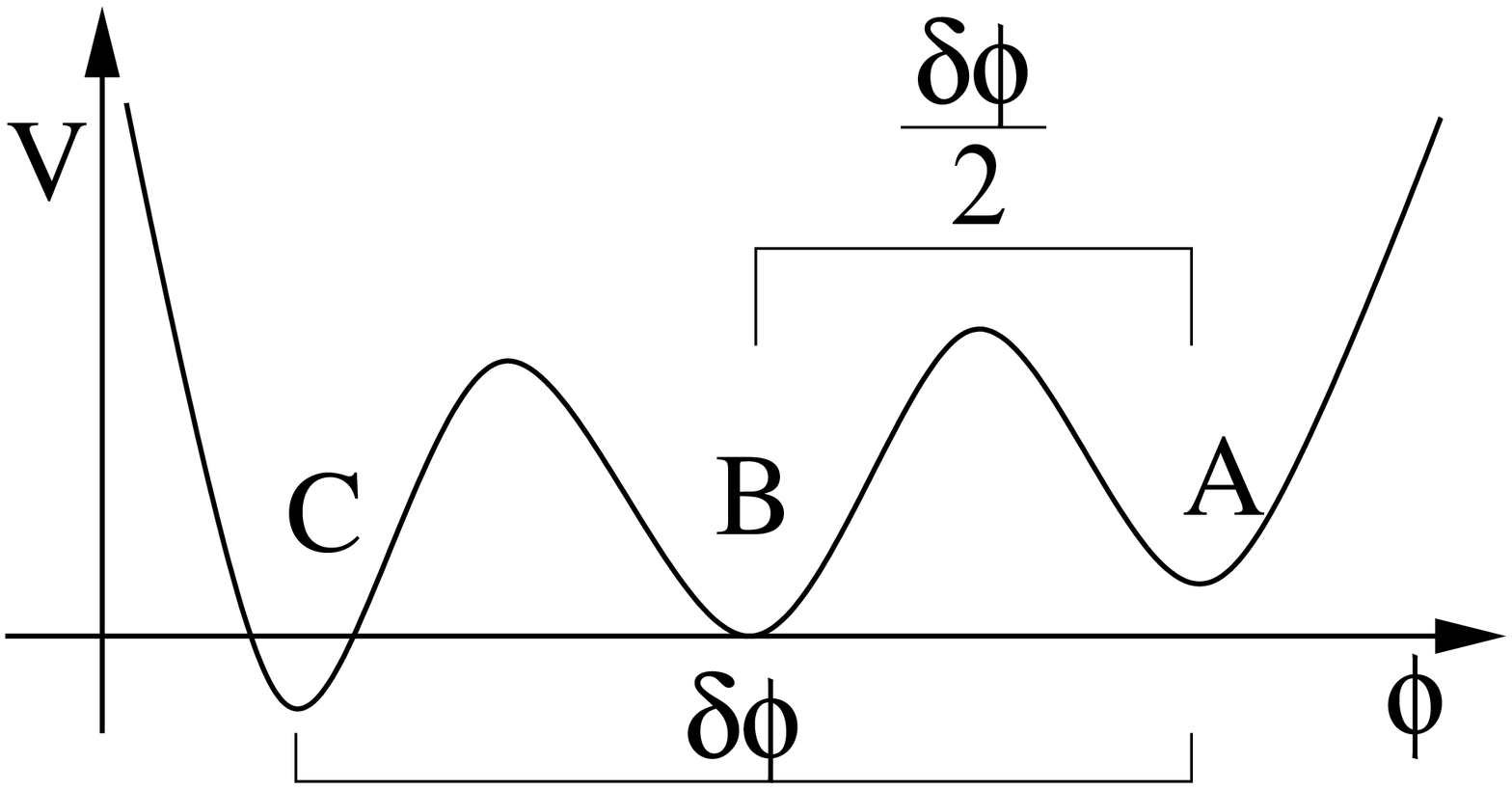}
\caption{Left: A classical transition.  Vacuum bubbles $B_1$ and $B_2$ collide and create a new vacuum region $C$. \\
Right: A scalar field potential that can give rise to a classical transition with $B_1=B_2=B$. Each of the three minima is separated by a distance $\delta \phi / 2$ in field space.}
\label{fig-CT}
\end{figure}

As shown in Fig.~\ref{fig-CT}, a classical transition refers to a collision that creates two new domain walls enclosing a region of some new vacuum. This was already observed in early simulations by Hawking, Stewart, and Moss~\cite{HawMos82a}, who studied a potential with two minima, and found that a collision can cause the field to jump back up to the false vacuum in a localized region. With a more complicated potential, containing more than two minima, the there are many more possibilities. 

\begin{figure}[tb]
   \includegraphics[width=7 cm]{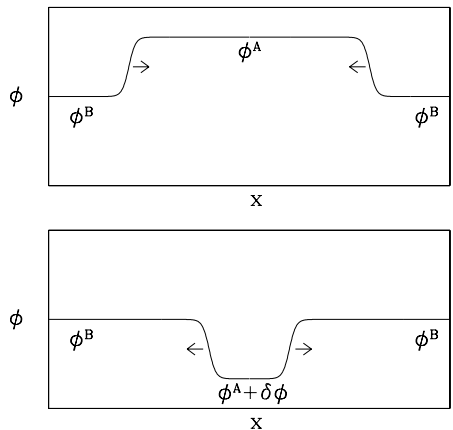} 
\caption{Field configurations before(top) and after(bottom) a classical 
transition.  The solitonic profiles superpose, and because of the equal spacing of the vacua in the potential, create a region in the $C$ phase which subsequently expands.}
\label{fig-freepass}
\end{figure}

The prototype of the classical transition we study is realized by the potential in Fig.~\ref{fig-CT}, which has three equally spaced vacua $A$, $B$, and $C$. Fig.~\ref{fig-freepass} shows a numerical simulation of the prototype classical transition. Surprisingly, when the domain walls 
separating vacuum $A$ from vacuum $B$ collide, the field profiles can stay compact. When the velocity of the walls is large enough, the  potential for the field becomes irrelevant as compared to the spatial gradients, and the solitonic field configurations simply superpose. Because the vacua are equally spaced, the domain walls now enclose a region of vacuum $C$, which subsequently expands due to its lower energy density as compared to vacuum $B$! It is also clear from the fact that the walls simply superpose that collisions can only cause excursions in field space of a fixed distance.

In a classical transition, the resultant field profile may not exactly be the soliton between $\phi_B$ and $\phi_C$, because the barriers are different. In particular, the height difference implies that the solitons have different mass (wall tension). In high resolution simulations~\cite{GibHui10}, it was observed that the profiles settle into the true solitonic configuration, in the process radiating very little energy. Instead, the velocity is adjusted to accommodate the mass change. The superposition condition, which comes from microphysics, is always more restrictive than energy conservation, although both give a lower bound on the incoming boost necessary to produce a classical transition.  

In general, the classical transition in Fig.~\ref{fig-CT} can be realized on a potential for multiple variables $\vec{\phi}$ if the following conditions are true.
\begin{itemize}
\item The incoming boost is large enough so that the solitonic profiles freely
pass through each other.
\item $\vec{\phi}_C=\vec{\phi}_{B_1}+\vec{\phi}_{B_2}-\vec{\phi}_A$ 
is a vacuum.
\item The solitonic profile between $A$ and $B_1$ can settle into the new profile separating $B_2$ and $C$.  Similarly $A$ and $B_2$ to $B_1$ and $C$.
\end{itemize}
The last two conditions may look finely tuned, but they can be easily arranged.~\footnote{In a toy model with fluxes and extra dimensions\cite{BlaJos09}, these conditions are guaranteed.} In the remainder of the paper, we will have little to add to the discussion of how ubiquitous classical transitions are in a realistic potential landscape.

\section{Classical transitions with gravity}\label{sec:junctionconditions}

Neglecting gravity, the vacuum energy on either side of a domain wall fully determines its dynamics. In a spacetime undergoing a classical transition, as depicted in Fig.~\ref{fig-CT}, the region of new vacuum (vacuum $C$) will expand indefinitely when its energy is lower than that of the colliding bubbles (vacuum $B$). If the region of new vacuum has higher energy, it is always doomed to collapse to zero size. Therefore, neglecting gravity, classical transitions can only seed lasting regions of new vacuum whose energy is lower than the energy inside of the colliding bubbles. 

How might the inclusion of gravity affect this picture? A full treatment of gravitational effects in a classical transition would require simulations employing numerical relativity. However, given a scalar potential with the appropriate vacuum structure, the field dynamics outlined in the previous section (more details will be presented in~\cite{GibHui10}) strongly suggest that it is reasonable to think of the classical transition spacetime as consisting of regions of vacuum separated by domain walls. We can therefore make an important simplification, and reduce the full field dynamics to a study of the motion of domain walls in a background spacetime. This simplified picture allows us to isolate several distinct phenomena: the gravitationally repulsive nature of domain walls~\cite{Ipser:1983db,Vilenkin:1984hy}, the effects of a non-zero cosmological constant on the motion of domain walls, and the gravitational backreaction of the energy released in the collision. 

As we will see, the presence of gravity alters the domain wall motion in a nontrivial way, modifying the energy conservation condition~\footnote{Energy conservation alone is not enough to guarantee a classical transition, since the conditions on the field dynamics can be much more stringent. In this paper, we will assume that a classical transition can proceed as long as the energy conservation is satisfied.}.  when the walls cross each other. However, the explicit form of energy conservation and the equations of motion is known and after working though this, the nontriviality in some cases conspires to allow for classical transitions to {\em expanding} regions with vacuum energy {\em higher} than that of the colliding bubbles. Such transitions are particularly dramatic when the colliding bubbles have negative vacuum energy, since the region of new vacuum can avoid the big-crunch singularity inside the bubbles, bringing a ``terminal" vacuum back to life.  

In the remainder of the paper, we will examine these features using the Israel Junction condition formalism. After describing the general setup and the relevant equations of motion and energy conservation in this section, we move on to study colliding bubbles with positive and negative vacuum energy in Sec.~\ref{sec-dS} and~\ref{sec-AdS} respectively.

\subsection{General setup}

The collision between two vacuum bubbles preserves an SO(2,1) symmetry~\cite{HawMos82a,Wu:1984eda}. Using a hyperbolic version of Birkhoff's theorem, the most general vacuum solution to Einstein's equations can be written as
\begin{equation}\label{eq:metric}
ds^2 = - a_I (z)^{-1} dz^2 + a_I (z) dx^2 +z^2 dH_2^2~,  
\end{equation}
where $dH_2^2 = d\chi^2 + \sinh^2 \chi d\phi^2$ is the metric on a unit hyperboloid, and the coordinates generally range from $0 < z < \infty$, $-\infty < x < \infty$, $-\infty < \chi < \infty$, and $0 < \phi < 2 \pi$. Allowing for vacuum energy, the metric functions $a_I(z)$ are 
\begin{equation}
a_I (z) = 1 - \frac{2 M_I}{z} + H_I^2 z^2~,
\end{equation}
where the label $I$ keeps track of different regions.  In our conventions, $H_I^2 > 0$ for positive vacuum energy, and $H_I^2 < 0$ for negative vacuum energy.

\begin{figure}
\includegraphics[width=12cm]{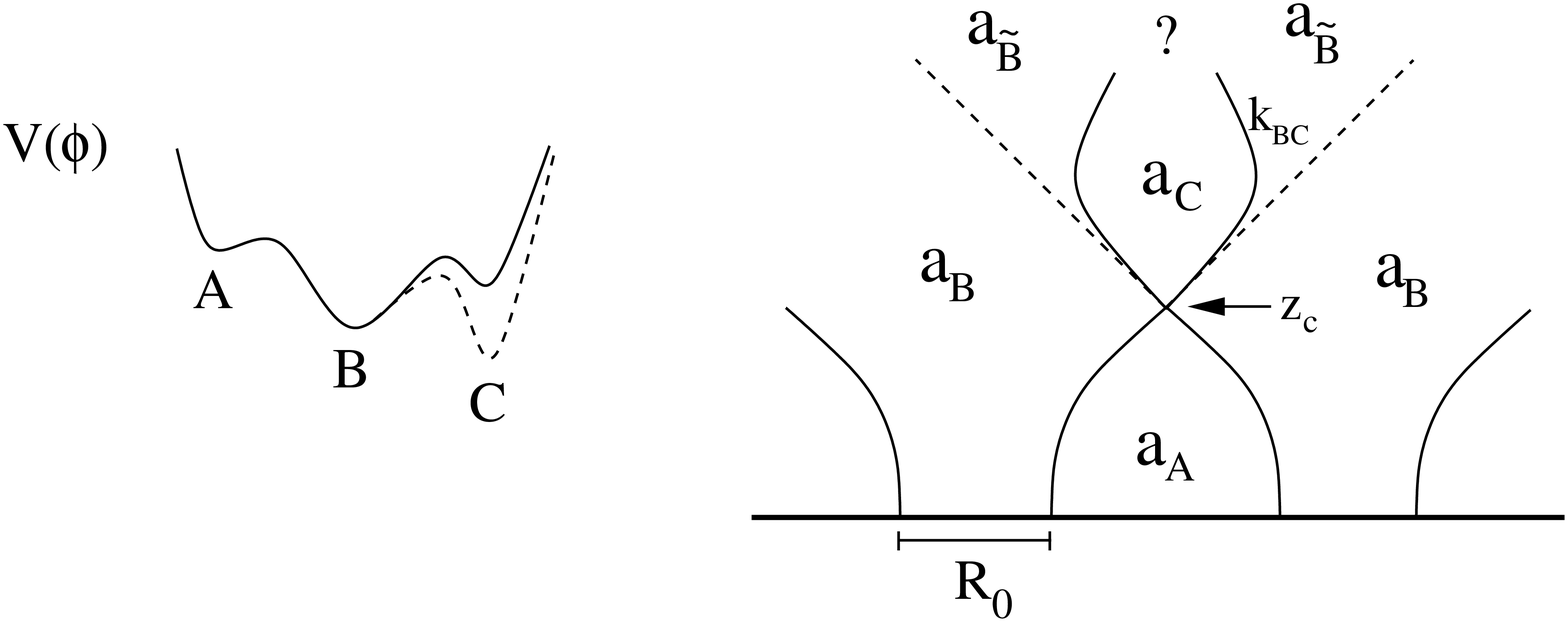}
\caption{The potential (left) and collision spacetime (right) that we consider for a classical transition. Two bubbles of the $B$ phase (of initial radius $R_0$) in a background of the $A$ phase collide at $z=z_c$, yielding a classical transition to vacuum $C$. The $C$ phase can have either higher or lower energy than the $B$ phase. There might be a small amount of additional debris released by the collision, which we account for as null shells, enclosing region $\tilde{B}$. Each region, described by the metric Eq.~(\ref{eq:metric}), is sewn together using the junction condition formalism. Our goal is to find the fate of region $C$. 
\label{fig:diagram}
}
\end{figure}

The setup for the collision problem is shown in Fig.~\ref{fig:diagram}, which depicts a slice in the $x,z$ plane. In the body of the text, we consider collisions between identical bubbles only, and briefly comment on the collision between different bubbles in Appendix~\ref{app-B1B2}. We assume that the underlying scalar potential has three vacua, $A$, $B$, and $C$ and can support a classical transition in the absence of gravity from vacuum $B$ to vacuum $C$. We take $H_A^2 > H_B^2$ so that true vacuum bubbles of the $B$ phase, with some initial radius $R_0 \ll H_{A}^{-1}$, can nucleate from the $A$ phase. When two bubbles of the $B$ phase collide at $z = z_c$ (in these coordinates, the collision occurs at a position specified by $z$), a classical transition can occur to the $C$ phase, which we allow to be of arbitrary energy $H_C^2$. A domain wall of normalized tension $k_{BC}$ (this quantity is related to the tension through the relation $k_{BC} = 4 \pi G \sigma_{BC}$) separates the $B$ and $C$ phases. In addition to the classical transition, the collision can release energy in other forms, which we model as null shells of radiation. We refer to the portion of the $B$ phase to the future of a radiation shell as the $\tilde{B}$ phase. While we generally include their effects, we neglect these energy sinks in many of the specific examples to follow.  

The metric in each of the four qualitatively different regions is of the form Eq.~(\ref{eq:metric}), with the metric coefficient in each region given by
\begin{equation}
a_A = 1 + H_A^2 z^2, \ \ a_B = 1 + H_B^2 z^2, \ \  a_{\tilde{B}} = 1 - \frac{2 M_{\tilde{B}} }{z} + H_B^2 z^2, \ \ a_C = 1 - \frac{2 M_C}{z} + H_C^2 z^2~. 
\end{equation}
There are two free parameters in the metric, $M_C$ and $M_{\tilde{B}}$, that are not determined by the underlying potential. 

The junction condition formalism gives a prescription for sewing each of these spacetime regions together in a consistent way. First, we require that the metric be continuous across all junctions. This implies that $z$ must be continuous throughout the collision spacetime. The derivatives of the metric can be discontinuous, and will be related to the energy density of local sources such as domain walls. In the collision spacetime, there are a number of separate interfaces to consider: the bubble walls between the $A$ and $B$ phases, the post-collision domain walls separating the $B$ and $C$ phases, and the radiation shells separating the $B$ and $\tilde{B}$ phases. The dynamics of each interface can be determined from Einstein's equations. Requiring that each of these interfaces are stitched together consistently is equivalent to imposing energy and momentum conservation at the location of the collision. Summarizing the necessary steps in solving for the collision spacetime, we must:
\begin{enumerate}
\item Solve for the trajectories of the in-coming bubble walls and the location of the collision. We work in a frame where the colliding bubbles are nucleated at the same time, and so the kinematics of the collision are fully determined by their initial radii and separation (see~\cite{Aguirre:2007wm} for a discussion of the various frames that can be used to describe bubble collisions). This yields the position $z=z_c$ of the collision.
\item At the collision, impose energy conservation to determine if a solution for $M_C$ and $M_{\tilde{B}}$ is possible for a set of input parameters: $H_A$, $H_B$, $H_C$, $z_c$, $R_0$, $k_{BC}$. If a solution exists, this provides the initial conditions for the domain walls enclosing the $C$ phase. 
\item Solve for the motion of the domain walls enclosing the $C$ phase using Einstein's equations.
\end{enumerate}

\subsection{Junction condition formalism}

A domain wall is a (2+1) dimensional boundary between two vacua, with metric
functions $a_L$ to the left and $a_R$ to the right. The geometry is continuous
but not smooth, so the $z$ coordinates are the same but $x_L$, $x_R$ can be 
different. The normal for the wall is defined to point from the L phase into the R phase, and requiring orthogonality, its components are given by 
\begin{equation}
{\rm  L:} \ \ n_z = \dot{x}_L, \ \ \ n_x = \dot{z}, \ \ \ {\rm R:} \ \ n_z = \dot{x}_R, \ \ \ n_x = \dot{z}~,
\end{equation}
where the dot refers to a derivative with respect to the proper time of an observer riding on the wall. Normalizing, we obtain the condition
\begin{equation}\label{eq:dotx}
\dot{x}_L = \pm \frac{\sqrt{\dot{z}^2 - a_L}}{a_L}, \ \ \ \dot{x}_R = \pm \frac{\sqrt{\dot{z}^2 - a_R}}{a_R}~.
\end{equation}

Integrating Einstein's equations across the wall, we obtain~\cite{Isr66},
\begin{equation}\label{eq:junction1}
K_{L \ b}^{a} - K_{R \ b}^{a} = -k_{LR} \delta_{b}^{a}~,
\end{equation}
where $K$ is the extrinsic curvature of the wall (which is different on either side),
\begin{equation}\label{extrinsic}
K_{ab} = \frac{1}{2} n^{\mu} \partial_{\mu} \gamma_{ab}~,
\end{equation}
$k_{LR} = 4 \pi G \sigma_{LR}$ is the normalized tension of the wall, and 
$a,b$ run through the (2+1) dimensions on the wall worldsheet. Although the components of Eq.~(\ref{eq:junction1}) appear to yield two independent equations, they are actually the same and simplify to
\begin{equation}
\beta_L-\beta_R = k_{LR} z~.
\label{eq-junc}
\end{equation}
Here $\beta_I = a_I \dot{x}_I$ is defined as the physical ``boost'' of the 
wall, which will simplify further calculations and their physical 
interpretations. 

From Eq.~\ref{eq-junc} and the definitions Eq.~(\ref{eq:dotx}), we get
\begin{eqnarray}
\dot{z}^2 &=&
\frac{1}{4}\bigg[z^2k_{LR}^2+2(a_L+a_R)+
\frac{(a_L-a_R)^2}{z^2k_{LR}^2}\bigg]~,  \nonumber \\
\dot{x}_L &=& \frac{\beta_L}{a_L}, \label{eq-z} \\
\dot{x}_R &=& \frac{\beta_R}{a_R}~, \nonumber
\end{eqnarray}
with
\begin{eqnarray}
\beta_L   &=& \frac{a_R-a_L+z^2k_{LR}^2}{2zk_{LR}}~, \label{eq-L} \\
\beta_R   &=& \frac{a_R-a_L-z^2k_{LR}^2}{2zk_{LR}}~. \label{eq-R}
\end{eqnarray}
In this form, we can already note an interesting property when $a_L$ and $a_R$ are positive definite. In this case, the sign of $\dot{x}_{L,~ R}$ are determined by the sign of $\beta_{L, ~ R}$. When $|a_R-a_L| > z^2k^2$, we have $\dot{x}_{L,~R} > 0$ or $\dot{x}_{L,~R} < 0$, and the wall moves away from the region with smaller $a_I$ into the region with larger $a_I$.  We will refer to this behavior as {\bf normal}. When $|a_R - a_L| < z^2k^2$, we have $\dot{x}_L > 0$ while $\dot{x}_{R} < 0$, and the wall moves away from both sides. We will refer to this as {\bf repulsive} behavior.  

Integrating Eq.~(\ref{eq-z}), it is possible to solve for the trajectory $x(z)$. Note that Eq.~(\ref{eq-z}) is just the equation of motion for a particle of unit mass with zero energy moving in an effective potential
\begin{equation}
V_{eff}=-\frac{1}{8}
\bigg[z^2k_{LR}^2+2(a_L+a_R)+\frac{(a_L-a_R)^2}{z^2k_{LR}^2}\bigg]~.
\label{eq:effpot}
\end{equation}
In general, one can consider initial conditions with either positive or negative $\dot{z}$. To better understand the motion in the effective potential, we can manipulate it in 2 ways:
\begin{eqnarray}
- 8 V_{eff}&=&
\frac{1}{k_{LR}^2 z^2} \left[z^2k_{LR}^2 + (a_L-a_R) \right]^2+4a_R , \\
&=&\frac{1}{k_{LR}^2 z^2} \left[z^2k_{LR}^2 + (a_R-a_L) \right]^2+4a_L~.
\label{eq-2tricks}
\end{eqnarray}
It can be seen that both $a_L < 0$ and $a_R < 0$ at some position $z$ are necessary conditions for a turning point in the motion of $z$ ($\dot{z} = 0$). Otherwise, the effective potential is negative definite, and the motion in $z$ must be monotonic from $z=z_c \rightarrow \infty$ or from $z = z_c \rightarrow 0$.

\subsection{Determining the kinematics of the collision}\label{sec:kinematics}

\begin{figure}
\includegraphics[width=8cm]{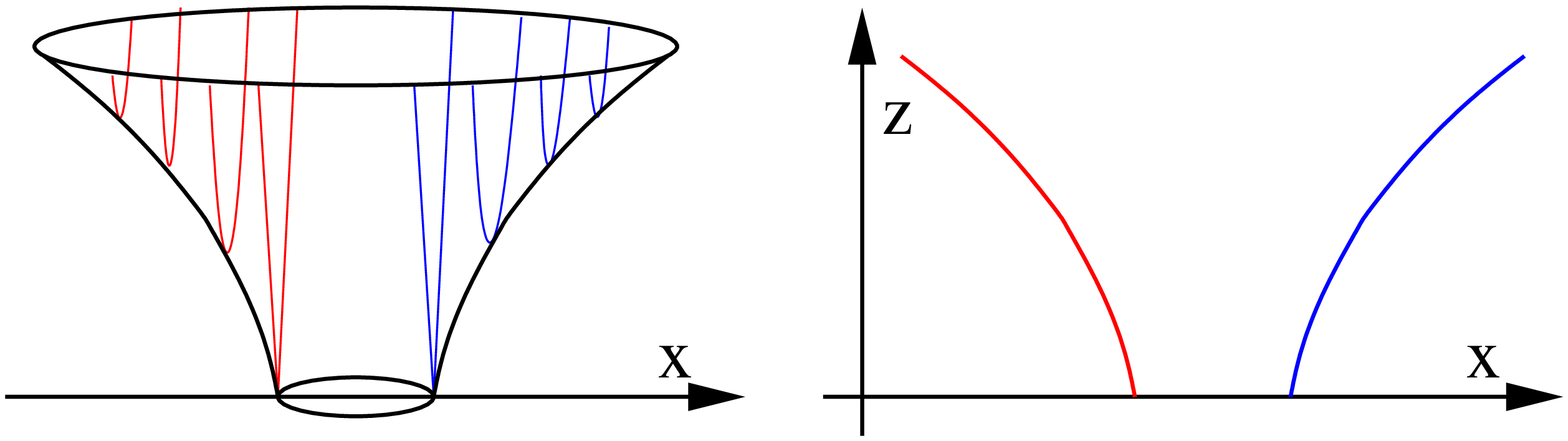}
\caption{Our coordinates only cover the interior of the forward lightcone of 
the $x$ axis.  Therefore, a full SO(3,1) symmetric bubble wall shows up as two 
separated portions, namely two separated lines on the $x-z$ picture.  Each
point there is a suppressed $H_2$ on the left picture.  Only the blue(right)
portion will be relevant to a collision with another bubble from a larger $x$
position.\label{fig-bubble}}
\end{figure}

Before the collision, the incoming domain wall in the SO(2,1) coordinates is only a portion of the SO(3,1) symmetric bubble 
wall, as shown in Fig.~\ref{fig-bubble}. By symmetry, we can choose the incoming bubble on the left to fully describe the initial conditions for the collision. With $a_L \rightarrow a_B=1+H_B^2z^2$, 
$a_R \rightarrow a_A=1+H_A^2z^2$, $k_{LR} \rightarrow k_{BA}$, 
the equations for the incoming domain wall are quite simple.
\begin{eqnarray}\label{eq:incomingwall}
\dot{z}_{\rm in}^2 &=& a_W = 1+\frac{z^2}{R_0^2}~, \\
\beta_{\rm in,B}   &=& \sqrt{a_W-a_B}~, \\
\beta_{\rm in,A}   &=& \sqrt{a_W-a_A}~,
\end{eqnarray}
where
\begin{equation}
R_0 = 2 k_{AB} \left[ \left( H_A^2 - H_B^2 \right)^2 + 
2 k_{AB}^2 \left( H_A^2 + H_B^2 \right) + k_{AB}^4 \right]^{-1/2}~.
\end{equation}
In order to collide, the incoming wall has to move into region $A$, which requires $\dot{x}_A > 0$. The wall trajectory always begins at $z = 0$ where $a_A >0$, and from Eq.~(\ref{eq-z}) we require $k_{BA}^2<H_A^2-H_B^2$ so that the wall is {\bf normal} and $R_0 < H_A^{-1}$.

If the bubble centers are initially separated by a distance $2 \Delta x$, then arranging the incoming walls so that they collide at $x=0$, the solution to Eq.~(\ref{eq:incomingwall}) for the trajectory of the wall is given by
\begin{equation}
z = H_A^{-1} \left[ \left( 1 - H_A^2 R_0^2 \right) \sec^2 \left[ H_A \left( x + \Delta x \right) \right] - 1 \right]^{1/2}~.
\end{equation}
if $H_A^2 > 0$ and
\begin{equation}
z = \left[ \left( x - \Delta x \right)^2 - R_0 \right]^{1/2}~,
\end{equation}
if $H_A = 0$. The location of the collision, $z=z_c$, is found by setting $x=0$ in these expressions.

\section{Colliding de Sitter and Minkowski bubbles}
\label{sec-dS}

In this section, we look for classical transitions in cases where the bubble interiors have $H_B^2 \geq 0$, leaving $H_C^2$ a free parameter. The collision occurs at $\{x=0, z=z_c \}$, and because the colliding bubbles are identical, there is a reflection symmetry about $x=0$. We can therefore choose to describe the collision by focusing on the incoming and outgoing walls on the left side. The motion of the incoming walls was found in Sec.~\ref{sec:kinematics}. After the collision, in the absence of null shells of radiation, the post-collision domain wall is described by
\begin{eqnarray}
\beta_{\rm out,B}  &=& \frac{a_C-a_B+z^2k_{BC}^2}{2zk_{BC}}~,  \\
\beta_{\rm out,C}  &=& \frac{a_C-a_B-z^2k_{BC}^2}{2zk_{BC}}~. 
\label{eq-betas}
\end{eqnarray}
In the $B$ region, the metric function $a_B$ is positive definite for $H_B^2 \geq 0$. In addition, by symmetry, we must have $a_C \geq 0$. Therefore, in order to have a region of $C$ that is initially growing after the collision ($\dot{x}_C < 0$), we must impose the physical constraint $\beta_C<0$. Since $a_B$ is positive definite, we have from Eq.~(\ref{eq-2tricks}) that after the collision $z$ must be monotonically increasing ($z$ is a timelike coordinate in $B$, and therefore initial conditions with $\dot{z}<0$ are not physical). 

The parameter $M_C$ in $a_C=1-2M_C/z+H_C^2z^2$ is determined by energy conservation. Technically, it says the boost angles around the collision point $z_c$, after moving across all crossing walls, must sum up to zero
\cite{Freivogel:2007fx,Aguirre:2007wm,Chang:2007eq}.
\begin{equation}\label{eq-boosts}
0 = \cosh^{-1} \left[ \frac{a_W^{1/2} }{ a_B^{1/2} } \right] - \cosh^{-1} \left[ \frac{a_W^{1/2} }{ a_A^{1/2} } \right] + \sinh^{-1} \left[ \frac{\beta_{\rm out,C}}{a_C^{1/2}} \right] - \sinh^{-1} \left[ \frac{\beta_{\rm out,B}}{a_B^{1/2}} \right]~.
\end{equation}
In the above equation, all functions of $z$ refer
to their specific values at the collision point $z_c$.  Also, since the 
incoming boosts are replaced by $a_W$, we will drop the ``out'' subscript
and $\beta$ from now on refers to the outgoing walls.  Further 
simplification leads to
\begin{equation}\label{eq-energy}
\frac{\sqrt{a_Aa_C}}{a_B}=
\bigg(\frac{\sqrt{a_W}+\sqrt{a_W-a_A}}{\sqrt{a_W}+\sqrt{a_W-a_B}}\bigg)
\bigg(\frac{-\beta_C+\sqrt{\beta_C^2+a_C}}
{-\beta_B+\sqrt{\beta_B^2+a_B}}\bigg)~.
\end{equation}

Although simulations ~\cite{Easther:2009ft,GibHui10} show that in single field models very little energy is radiated in a classical transition, the addition of extra degrees of freedom might allow for significant energy loss\cite{Zhang:2010qg}. A detailed picture of how much energy is lost and in what form is beyond the thin-wall analysis that we perform. However, the main impact is to make a classical transition harder by taking some energy away. In an attempt to model this, we include shells of radiation as shown in Fig.\ref{fig:diagram}. This choice of energy sink is motivated by simulations~\cite{HawMos82a,Aguirre:2008wy}. In this case, the additional region has the metric function
\begin{equation}
a_{\tilde{B}} = 1 - \frac{2M_{\tilde{B}}}{z}+H_B^2z^2~,
\end{equation}
where $M_{\tilde{B}}$ is related to the energy density of the radiation shells by
\begin{equation}
\sigma_{\rm rad} = \frac{M_{\tilde{B}}}{4 \pi z_c^2}~.
\end{equation}
The boost of outgoing walls takes exactly the forms in Eq.~(\ref{eq:betas})
with $a_B$ replaced by $a_{\tilde{B}}$.  
\begin{eqnarray}
\beta_{\tilde{B}}  &=& \frac{a_C-a_{\tilde{B}}+z^2k_{BC}^2}{2zk_{BC}}~,  \\
\beta_{C}  &=& \frac{a_C-a_{\tilde{B}}-z^2k_{BC}^2}{2zk_{BC}}~.
\label{eq:betas}
\end{eqnarray}
The energy conservation equation picks up an additional term,
\begin{equation}\label{eq:boostcondition}
0 = \cosh^{-1} \left[ \frac{a_W^{1/2} }{ a_B^{1/2} } \right] - \cosh^{-1} \left[ \frac{a_W^{1/2} }{ a_A^{1/2} } \right] + \sinh^{-1} \left[ \frac{\beta_C}{a_C^{1/2}} \right] - \sinh^{-1} \left[ \frac{\beta_{\tilde{B}}}{a_B^{1/2}} \right] + \frac{1}{2} \log \left[ \frac{a_{\tilde{B}}}{a_B} \right]~,
\end{equation}
which in the end is very similar to Eq.~(\ref{eq-energy}).
\begin{equation}\label{eq:energyconservation}
\frac{\sqrt{a_Aa_C}}{a_B}=
\bigg(\frac{\sqrt{a_W}+\sqrt{a_W-a_A}}{\sqrt{a_W}+\sqrt{a_W-a_B}}\bigg)
\bigg(\frac{-\beta_C+\sqrt{\beta_C^2+a_C}}
{-\beta_{\tilde{B}}+\sqrt{\beta_{\tilde{B}}^2+a_{\tilde{B}}}}\bigg)~.
\end{equation}
Note that with the extra variable $M_{\tilde{B}}$, there are not enough constraints to fully determine the geometry, and we need detailed information about the microphysics as discussed above. In practice, we will treat $M_{\tilde{B}}$ a tunable input parameter.

\subsection{Null limit}
\label{sec-null}
There is a particularly simple form of the energy conservation equations when the incoming and outgoing domain walls are highly boosted. The incoming walls will be highly boosted when they are light ($k_{AB}^2\ll H_A^2-H_B^2$) and $z$ is large ($z_c\gg H_A^{-1}$). In this case, we have that  $a_W\gg a_A$, $a_W\gg a_B$. The out going walls will be highly boosted when $-\beta_C^2\gg a_C$ and $-\beta_B^2 \gg a_B$~ (or $-\beta_{\tilde{B}}^2\gg a_{\tilde{B}}$ if radiation shells are included). In this limit, both Eq.~(\ref{eq-energy}) and Eq.~(\ref{eq:energyconservation}) simplify to
\begin{equation}
a_A a_C = a_B^2~.
\label{eq-null}
\end{equation}
This is exactly true for null domain walls and numerically verified to be a 
good approximation for highly relativistic walls.  We can use this simple 
relation to examine the general properties of classical transitions.

In this limit, $a_C (z_c)$ is independent of $H_C^2$, which is apparent after solving for the value of $M_C$:
\begin{equation}
1-\frac{2M_C}{z_c}+H_C^2z_c^2 = \frac{(1+H_B^2z_c^2)^2}{1+H_A^2z_c^2}~.
\end{equation}
For $H_B^2 < H_A^2$, we always have $a_C<a_B$ at the 
collision point, and the domain wall is {\bf normal}, moving from region $C$
into region $B$. This makes sense as the collision is a local event, which 
only concerns the incoming and outgoing walls, but not the properties of the spacetime
regions in between (such as their vacuum energy).

However, as $z$ increases at late times, the $M_C$ term in $a_C$ becomes irrelevant and only 
the vacuum energy matters. In this regime, the signs of $\dot{x}_B$ and $\dot{x}_C$ depend only on the 
 values of the energy density in either vacuum and the tension of the post-collision domain wall:
 \begin{eqnarray}
{\rm sign} (\dot{x}_B) \rightarrow {\rm sign} (H_C^2 - H_B^2 + k_{BC}^2)~, \\
{\rm sign} (\dot{x}_C) \rightarrow {\rm sign} (H_C^2 - H_B^2 - k_{BC}^2)~.
\end{eqnarray}
Thus, adjusting $H_C^2$ controls the asymptotic behavior of the outgoing walls, and produces a complete list of possible geometries as the results of classical transitions from identical bubble collisions:

\begin{itemize}
\item $H_B^2-H_C^2>k_{BC}^2$~{\it Normal} \\
      After the collision, the domain walls always move from region $C$ into
      region $B$ ($\dot{x}_{C} < 0$, $\dot{x}_{B} < 0$) . Namely, they are always {\bf normal}. 
\item $|H_B^2-H_C^2|<k_{BC}^2$~{\it Repulsive}  \\
      The initially {\bf normal} domain walls eventually become 
      {\bf repulsive}.  As a result, an expanding region $C$ seems to be moving
      away from region $B$ on both sides of the wall ($\dot{x}_{C} < 0$, $\dot{x}_{B} > 0$) .
\item $H_C^2-H_B^2 \gg k_{BC}^2$~{\it Oscillatory} \\
      Immediately after the collision, the domain walls move from region $C$
      into region $B$, but eventually turn back and collide with each other.
      The second collision can induce another transition from $C$ back to $A$ 
      with similar domain wall motion, resulting in an oscillating series of 
      transitions. In Appendix~\ref{app-oscillate} we outline the properties 
      of these geometries in more detail, showing that each successive 
      collision is less and less energetic.  This loss of energy can cause the 
      oscillations to terminate, at which point the collision energy must be 
      dissipated in some other form. If the oscillations last long enough, the 
      oscillating region acts as a Repulsive domain wall separating the 
      interiors of the two colliding bubbles.      
\item $H_C^2-H_B^2 \gtrsim k_{BC}^2$~{\it Marginally Repulsive}   \\
      A while after the collision, the initially expanding domain walls turn 
      back to move from region $B$ into region $C$ ($\dot{x}_{C} > 0$, $\dot{x}_{B} > 0$). However, in classical 
      transitions with $H_C^2 > 0$, the expansion in region $C$ and the 
      gravitationally repulsive nature of domain walls prevents the walls 
      from colliding again.
\end{itemize}

\begin{figure}
\includegraphics[width=16cm]{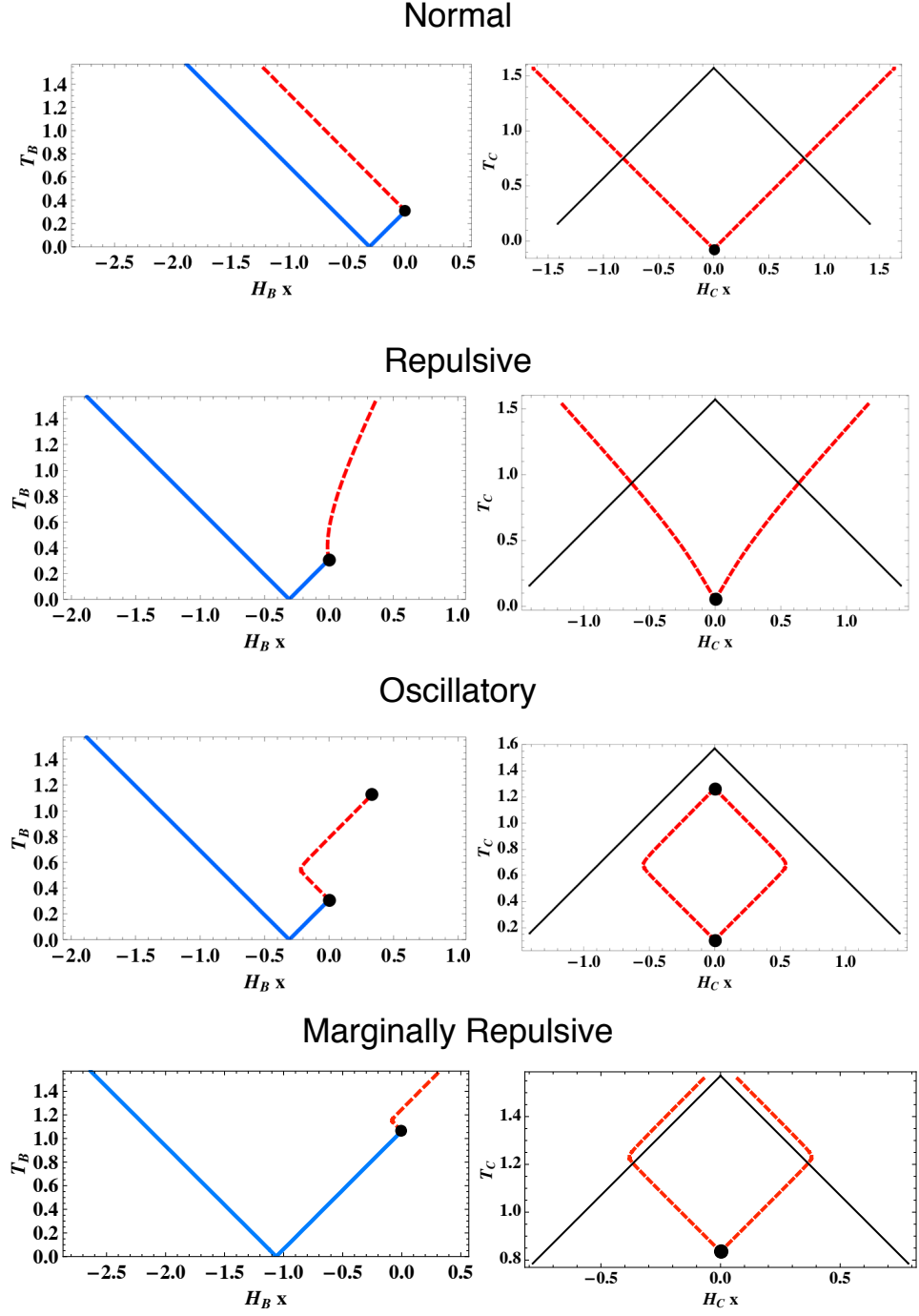}
\caption{Examples of the Normal, Repulsive, Oscillatory, and Marginally Repulsive classical transition geometries.  The left figure is the $B$ bubble on the left side of the collision, with its initial domain wall to vacuum $A$(blue, solid) and the domain wall to vacuum $C$(red, dotted) after the collision.  The right figure is the middle region $C$ with its domain walls(red, dotted) and the event horizon(black, solid).  In all but the Oscillatory geometry, an infinite region of the $C$ vacuum is produced by the classical transition. In each case, we choose the parameters (in units of $H_A$) Normal: $\{R_0 = .01, \Delta x = 2.9, H_C=.3, H_B=.2, k=.01 \}$, Repulsive: $\{R_0 = .01, \Delta x = 2., H_C = .4, H_B = .2, k = .4 \}$, Oscillatory: $\{R_0 = .01, \Delta x = 2., H_C=.3, H_B=.2, k=.01 \}$, and Marginally Repulsive: $\{R_0 = .01, \Delta \eta = 2.9, H_C=.3, H_B=.2, k=.01 \}$.
\label{fig:trajexamples}
}
\end{figure}

Finding exact solutions to the energy conservation equation Eq.~(\ref{eq-energy}), we then solve the equations of motion for $x$ and $z$  (Eq.~\ref{eq-z}) numerically to confirm the existence of each of these geometries. Such numerical simulations are key to determining if and when the Marginally Repulsive and Oscillatory geometries exist, since they depend on the details of the evolution away from the collision. Examples with $H_B^2 > 0$ and $H_C^2 > 0$ are shown in Fig.~\ref{fig:trajexamples}. In this series of plots, we show the wall trajectory as a function of the conformal time
\begin{equation}
H_I T_I = \int dz \frac{1}{a_I}~.
\end{equation}
In these coordinates, future infinity is located at $H_I T_I = \pi / 2 $. The location of a collision is denoted by the solid disc, and the post-collision walls correspond to the red dashed line. The interior of the second colliding bubble is a mirror image of the plots on the left.

The Oscillatory solutions are described in detail in Appendix~\ref{app-oscillate}, and here we show the first transition to the region of $C$. The field dynamics suggest that the secondary collision will yield a classical transition back to $A$. For the type of potentials we consider, the oscillations will eventually terminate, and the bubbles of $B$ will merge. The Marginally Repulsive geometries are more interesting for the question of which vacua can be populated in a classical transition. In our numerical simulations, the existence of these geometries can be easily verified by first drawing the past light cone from the point $x=0$ at future infinity in the region of $C$. If the wall exits this light cone, it can never recollapse to zero size. Therefore, even though $\dot{x}_C > 0$, the region of $C$ will continue to grow in physical volume. We have verified the existence of these geometries over a wide range of parameters, and we show a representative example in Fig.~\ref{fig:trajexamples}. 

In the absence of gravity, only the Normal and Oscillatory geometries exist, and a lasting region of the $C$ vacuum can only be produced if $H_C^2 < H_B^2$. Including gravity, the Repulsive and Marginally Repulsive geometries allow for lasting regions of $C$ even when $H_C^2 > H_B^2$! This is due to a combination of the gravitationally repulsive nature of domain walls and the background expansion of the $C$ vacuum, and is a key result of this paper. We now discuss the range of parameters in it is possible to find this behavior. 

\subsection{Constraints on possible geometries}
\label{sec-constraints}

Here we will go beyond the null approximation to investigate parameter 
constraints on the possible geometries, especially for the (marginally) 
Repulsive geometry.  The constraints come from simple physical intuition---energy 
conservation prevents the creation of arbitrarily heavy objects.  Although the 
incoming domain walls can accelerate for an unbounded amount of time, the 
expansion of de~Sitter vacuum $A$ still makes the total energy (in the center 
of mass frame) bounded. We should therefore expect a bound on the tension $k_{BC}$ 
of the post-collision domain walls that a collision is able to produce. Let us now see how this arises.

\begin{figure}
\includegraphics[width=8cm]{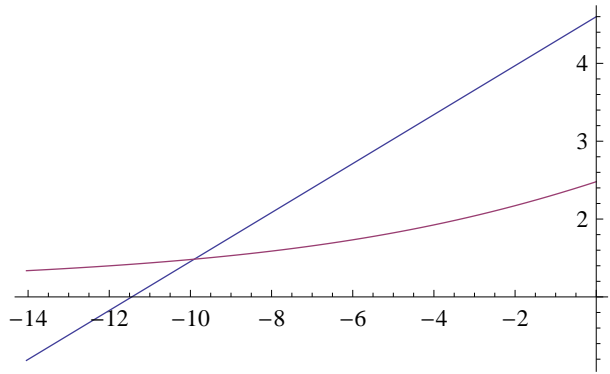}
\caption{The blue curve is left hand side of Eq.~(\ref{eq-betaC}), which goes
to zero at some finite $\beta_C$.  The red curve is the right hand side, which 
asymptotes to 1 as $\beta_C\rightarrow-\infty$.  The parameters here are 
$H_A=2$, $H_B=1$, $H_C=0.2$, $k_{BC}=0.4$, $z_c=10$, $R_0=0.1$, 
$M_{\tilde{B}}=0$.
\label{fig:betaC}
}
\end{figure}

For simplicity, we still demand the vacuum bubbles of $B$ to have small 
critical size, $R_0\ll H_A^{-1}$, therefore the second bracket on the right hand side of Eq.~(\ref{eq:energyconservation}) is approximately 1. Using the relation $\beta_B = \beta_C + k_{BC}^2 z^2$, we obtain
\begin{equation}
\frac{\sqrt{a_A(a_{\tilde{B}}+z_c^2k_{BC}^2+2z_ck_{BC}\beta_C)}}{a_B}
=\frac{-\beta_C+\sqrt{a_C+\beta_C^2}}
{-\beta_C-z_ck_{BC}+\sqrt{a_{\tilde{B}}+(\beta_C+z_ck_{BC})^2}}~.
\label{eq-betaC}
\end{equation}
The physical range that $\beta_C$ can take goes from $\beta_C = 0$, in which case the post-collision domain walls are produced at rest, to $\beta_C = -\infty$, where the walls are highly relativistic.
Plotting both sides of this equation as functions of $\beta_C$ in the physically allowed range, $\beta_C<0$, we get Fig.\ref{fig:betaC}. The two curves have to cross somewhere in this range in order to 
have a solution. As we can see, the left hand side goes to zero at some finite
$\beta_C$ while the right hand side asymptotes to 1 as 
$\beta_C\rightarrow-\infty$.  Therefore, the existence of a solution implies
that at $\beta_C=0$, the left hand side is larger than the right hand side.
\begin{equation}
\frac{\sqrt{a_A}}{a_B}>\frac{1}
{-z_ck_{BC}+\sqrt{a_{\tilde{B}}+z_c^2k_{BC}^2}}~.
\end{equation}
After several lines of algebra, we get
\begin{equation}
k_{BC}<\frac{a_{\tilde{B}}a_A-a_B^2}{2z_ca_B\sqrt{a_A}}~.
\label{eq-kbound}
\end{equation}
This agrees perfectly with our intuition:
\begin{itemize}
\item Setting $\beta_C$ to zero saturates an upper bound on $k_{BC}$. 
      That is, the heaviest possible out going walls are produced at rest.
\item Energy leaks into shells of radiation imply a positive $M_{\tilde{B}}$, 
      therefore a smaller $a_{\tilde{B}}<a_B$, which makes the upper bound for
      $k_{BC}$ smaller. The addition of energy sinks makes classical transitions harder.
\item As $z_c$ increases, the bound on the tension gets larger but asymptotes to 
      \begin{equation}
      {\rm Max}(k_{BC})=\frac{H_A^2-H_B^2}{2H_A}~.
      \label{eq-kmax}
      \end{equation}
      The center of mass energy is bounded for $H_A^2 > 0$.
\end{itemize}

Now that we have specific constraints on the possible classical transition geometries, we can return to the question of when a classical transition can produce a lasting region of vacuum $C$. Again, we are particularly interested in cases where $H_C > H_B$, since gravitational effects will be important in this regime. One way to guarantee a surviving region of $C$ is to produce a Repulsive geometry, which requires that $k_{BC}>H_C^2-H_B^2$. Combining this with the maximum tension Eq.~(\ref{eq-kmax}), we obtain a bound on the vacuum energy of the $C$:
\begin{equation}\label{eq:boundHC}
H_C^2 < \frac{\left( H_A^2 + H_B^2 \right)^2}{4 H_A^2}~.
\end{equation}
For vacuum energies satisfying this bound, energy conservation will allow a Repulsive geometry. Note that by default $H_B^2<H_A^2$ (which is required for bubbles expand and collide), so this also implies $H_C^2<H_A^2$. Numerical study shows that the Marginally Repulsive geometry also can only exist for $H_C^2<H_A^2$. Therefore, it is possible to produce a classical transition to lasting regions of $C$ with $H_C^2 > H_B^2$ only when $H_C^2 < H_A^2$. 

\subsection{Embedding regions of higher energy density}

Naively, it might seem surprising that it is possible to find geometries where a lasting region of $C$ can be produced with $H_C^2 > H_B^2$. Viewed from the perspective of vacuum $B$, this is a false vacuum (high energy density) ``blob." In spherically symmetric geometries, there is a theorem due to Vachaspati and Trodden (VT)~\cite{Vachaspati:1998dy} showing that a false vacuum region must be larger than the true vacuum horizon size. At first, our solutions might seem to be in conflict with this theorem: the false vacuum region $C$ grows from zero size, and then inflates indefinitely.  It is not an exact logical conflict as our geometry lacks spherical symmetry, but that is a meek point. Spherical symmetry was only a technical convenience used to bypass the complication of caustics in the theorem. We would like to provide a better physical reason to reconcile the apparent paradox.

Let us examine the assumptions of the VT theorem. First, the Null Energy Condition (NEC) must be satisfied. VT additionally assume spherical symmetry, and then the Raychaudhuri equation for the divergence of a spherical congruence of null geodesics can be written as
\begin{equation}
\frac{d \theta}{dT} \geq 0~.
\end{equation}
The content of this equation is that initially converging in-going null rays cannot diverge unless they pass through the origin. Said differently, following a null ray, the surfaces it passes through cannot go from being normal to anti-trapped. For a spherically-symmetric region of false vacuum embedded in some background with $H_T^2 > 0$, this condition implies that the initial size of the false vacuum region must be larger than $H_T^{-1}$. 

\begin{figure}
\includegraphics[width=10cm]{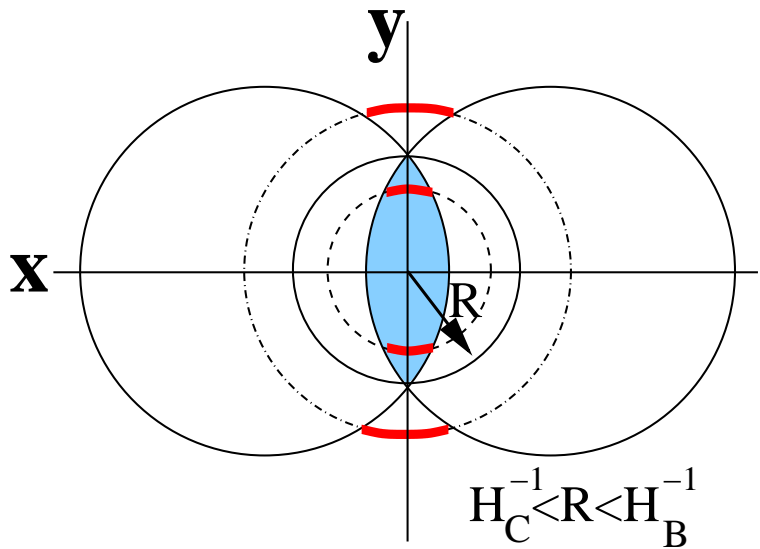}
\caption{The classical transition geometry on a constant-time slice. The region in the $C$ phase is shaded (blue). A surface of radius $H_{C}^{-1} < R < H_{B}^{-1} $ that encloses the region of $C$ is shown as the solid circle. A spherical congruence of ingoing null rays an instant before and after this time slice are depicted as the dot-dashed and dashed lines respectively. Diverging portions of the congruence are denoted by the solid red lines.
 \label{fig:diagram2}
}
\end{figure}

The bubble collision geometry we are considering is shown in Fig~\ref{fig:diagram2}, which depicts a constant time slice through the classical transition. Consider the case where the tensions are small, and the vacuum energies are comparable in magnitude. We can then define a set of flat coordinates, approximately given by the flat slicing of the exterior de Sitter space, whose origin is located at the center of the region of $C$. In these coordinates, the physical volume removed from the $B$ phase is equal to the physical volume replaced by the $C$ phase. Now, draw a closed spherical surface, whose radius is smaller than $H_B^{-1}$, but larger than $H_C^{-1}$, that completely encloses the region of $C$ at some time. Following a congruence of null rays inward from this surface, rays in the $B$ phase will be converging, while rays in the $C$ phase will be diverging. The shear and twist induced across the various interfaces will be minimal as long as the vacuum energies are comparable and tensions small. Now, tracing a congruence of null rays that move inward {\em towards} this surface, null rays in the $B$ phase will again be converging, but as long as $H_A > H_C$, the null rays in the $A$ phase will be diverging. The surfaces that the congruence of null rays pass through can therefore always be anti-trapped as long as $H_A > H_C$, and there is no conflict with the VT theorem. The reason it is possible to embed a region that is higher energy than the $B$ phase is that in the collision geometries, the $A$ and $C$ phases are always in contact.

There is another, related, theorem put forward by Guth and Farhi~\cite{FarGut87}: embedding a small and growing region of false vacuum in asymptotic Minkowski space must be accompanied by an initial singularity. This is an application of a singularity theorem due to Penrose~\cite{Penrose:1964wq}, and does not require spherical symmetry.  It does however require the existence of a noncompact Cauchy surface---a condition our geometry naively violates by having a de~Sitter parent vacuum with $H_A^2>0$. However, Farhi and Guth made a reasonable conjecture, based on the fact that local physics should not care about the cosmological scale $H_A^{-1}$. They claimed that the same theorem holds for de~Sitter spaces as well, if all our operations to generate the false vacuum region are ``local''. Indeed, it can be demonstrated that regions of false vacuum embedded in a background de Sitter space either require the existence of an initial singularity, or must be larger than the {\em exterior} horizon size~\cite{Aguirre:2005nt}. Our geometry pushes the  conjecture of Guth and Farhi to the extreme, and verifies their claim. Since the colliding bubbles can be almost $2H_A^{-1}$ apart, the initial configuration of the two colliding bubbles can be considered ``local'' only in the sense that the bubble walls eventually collide, and produce a classical transition, in the asymptotic future.

\section{Crunching bubbles}
\label{sec-AdS}

In this section, we focus on the case where $H_A^2 \geq 0$ and $H_B^2 < 0$. The interior of bubbles with 
$H_B^2 < 0$ undergo a big-crunch~\cite{Abbott:1985kr}, which has earned minima with negative energy the colorful title of ``terminal" vacua -- this is supposedly the end of the line for any observers inside of the bubble. But is it? We have seen that classical transitions can produce a lasting, non-singular, inflating region of $C$. Could a classical transition allow some portion of the bubble interior to avoid the crunch? At first glance, the prognosis is encouraging. For example, by taking $H_A = 0$, we can have arbitrarily large center of mass energy in the collision, and the bound on the tension $k_{BC}$ in Eq.~(\ref{eq-kmax}) does not exist. This implies that it might {\em always} be possible to create heavy walls, and therefore Repulsive geometries, in which case the region of $C$, regardless of its vacuum energy, is guaranteed to survive.

\begin{figure*}
\begin{center}
\includegraphics[scale = 0.5, angle=0]{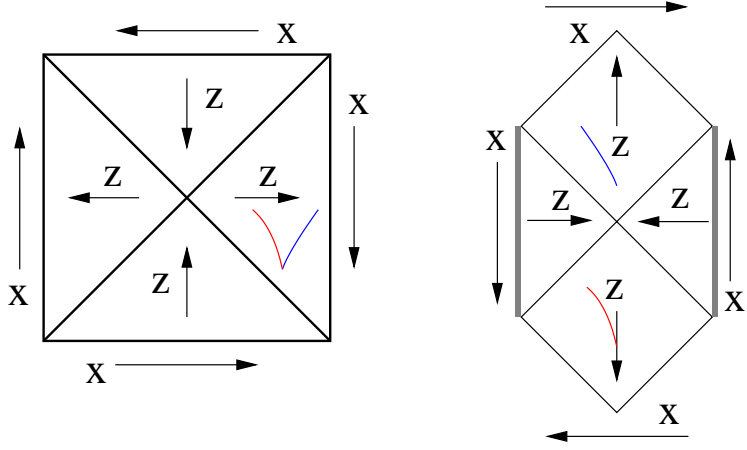}
\end{center}
\caption{The Penrose diagrams for AdS(left) and Hyperbolic-Schwarschild (right), for $M_C > 0$. The $x$ and $z$ coordinates are increasing in the directions labeled by the arrows. The blue curve (left-moving in the AdS diagram) has $\dot{z}>0$ and the red curve (right-moving) has $\dot{z}<0$, but they have the same $\beta_B = a_B \dot{x}_B > 0$.}
\label{fig-AdS}
\end{figure*}

However, producing an inflating region of $C$ from the collision of bubbles in Minkowski space is in conflict with our bound $H_C^2 < H_A^2$ from the previous section. One technical twist in the case of AdS bubbles is that there are horizons, across which $z$ becomes the spacelike coordinate and $x$ becomes timelike. 
See Fig.\ref{fig-AdS} for a complete coordinate cover. For $z < H_B^{-1}$, the 
matching problem is exactly the same as $H_B^2\geq0$. Highly 
boosted domain walls will collide in the region where $z_c > H_B^{-1}$. 
Something interesting happens there, and we will see that the bound $H_C^2 < H_A^2$ on surviving regions of $C$ is enforced not through kinematics, but through the large-scale structure of the classical transition spacetime.

In Sec.~\ref{sec-matching} we will first demonstrate how the different matchings 
work and how they depend on various parameters. Then in Sec.~\ref{sec-escapeAdS} we
will discuss the consequences: It is possible to make a lasting inflating region $C$ 
in the collision of two AdS bubbles, allowing some regions to escape from the crunch. However,
such regions cannot have a higher vacuum energy than the parent vacuum $A$, 
just as we found before in the collision of de Sitter and Minkowski bubbles.

\subsection{General Matchings}
\label{sec-matching}

We start from an example with $H_C^2=0$, which will capture most of the physics we want to understand. The formulas we derive, however, will be completely general. 
The metric in region $C$ is Hyperbolic-Schwarschild, with metric function $a_C=1-\frac{2M_C}{z}$. The causal structure is drawn in Fig.~\ref{fig-AdS}. 

In the region where $z^2 H_B^2 > 1$, $\beta_B$ is guaranteed to be positive, since $a_B < 0$ and $x$ is a monotonically decreasing timelike coordinate. There will be two timelike vectors with the same $\beta_B$, one with positive $\dot{z}$ the other negative.  This is not an ambiguity, but really two physically distinct situations.  When $\dot{z}>0$, the collision in region $C$ is located in the expanding quadrant; when $\dot{z}<0$, the collision is in the shrinking quadrant.

In terms of equations, the $\dot{z}>0$ case is simpler and is where we shall 
start. The analysis is nearly identical to the dS and Minkowski matchings we studied
earlier, but for convenience we will rearrange the energy conservation 
Eq.~(\ref{eq-energy}) to
\begin{equation}\label{eq:dotzg0energy}
\frac{\sqrt{a_Aa_C}}{-a_B}=
\bigg(\frac{\sqrt{a_W}+\sqrt{a_W-a_A}}{\sqrt{a_W}+\sqrt{a_W-a_B}}\bigg)
\bigg(\frac{-\beta_C+\sqrt{\beta_C^2+a_C}}
{\beta_B-\sqrt{\beta_B^2+a_B}}\bigg)~,
\end{equation}
so that both sides are positive definite. Unlike the collision between dS and Minkowski bubbles, there is no null limit of this equation where $\beta_C \rightarrow - \infty$. This is due to the fact that $x_B$ is a {\em timelike} coordinate for $z_c > |H_B|^{-1}$, which from Fig.~\ref{fig-AdS}, implies that $\beta_B > 0$. The condition $\beta_B = \beta_C + k_{BC} z$ then shows that $\beta_C$ cannot become arbitrarily negative.

Following the analysis of Eq.~(\ref{eq-betaC}), by looking at Eq.~(\ref{eq:dotzg0energy}) in the 
physical range $\beta_C<0$ and $a_C=0$, we can derive bounds on $k_{BC}$.  Although
the starting equation looks identical, the new roll of $x_B$ as a timelike
coordinate yields an interesting twist: comparing the two sides of the equation, in addition to the upper bound, we also
get a lower bound on $k_{BC}$: 
\begin{equation}
\frac{a_A+(-a_B)}{2\sqrt{a_A}}>z_ck_{BC}>\frac{-a_B}{\sqrt{a_A}}~.
\label{eq-rangek}
\end{equation}
For this range to exist, we need $(-a_B) < a_A$.
Note that in order to guarantee $x_B$ timelike, $k_{BC}$ is already bounded
\begin{equation}
z_c^2k_{BC}^2>a_C-a_B>-a_B~.
\end{equation}
Under the condition that $(-a_B)<a_A$, this lower bound is strictly better than
Eq.~(\ref{eq-rangek}), so logically we should replace it.
\begin{equation}
\frac{a_A+(-a_B)}{2\sqrt{a_A}}>z_ck_{BC}>\sqrt{a_C-a_B}~.
\end{equation}
This lower bound depends on a dynamical variable $a_C$, therefore cannot truly
be part of a physical constraint. It will soon become obvious that this bound 
comes from our assumption of $\dot{z}>0$, while nothing spectacular happens 
when $\dot{z}$ changes sign.

For $\dot{z}<0$, as can be seen in Fig.\ref{fig-AdS}, the solution should have $\beta_C>0$ because $x_C$ is increasing as one moves out of the interior of the region of $C$. The corresponding equation energy conservation is in this case slightly different from Eq.~(\ref{eq-energy}) (the first factor from incoming walls is already set to one).
\begin{equation}
\frac{\sqrt{a_Aa_C}}{-a_B}=\frac{\sqrt{\beta_C^2+a_C}+\beta_C}
{\sqrt{\beta_B^2+a_B}+\beta_B}~.
\label{eq-weird}
\end{equation} 
The first good sign here is that the null limit exists, and it is exactly $a_Aa_C=a_B^2$ as before. For a physical constraint on $k_{BC}$, we again compare both sides of Eq.~(\ref{eq-weird}) as functions of $\beta_C$ over its
physical range $\beta_C >0$. When $\beta_C\rightarrow \infty$, the right hand side of Eq.~(\ref{eq-weird}) goes to one, while the left hand side goes to infinity. A solution is guaranteed if there exists some value of  $\beta_C > 0$ where the right hand side is greater than the left hand side. The strictest bound on the tension comes when Eq.~(\ref{eq-weird}) is evaluated at some physical lower-bound on $\beta_C$. There is always a solution if $k_{BC}^2z_c^2<(-a_B)$, since the physical lower bound is $a_C=0$ in this case. No lower bound on the tension exists for solutions with $\dot{z} < 0$. If $a_C-a_B>k_{BC}^2z_c^2>-a_B$, then the physical lower bound on $\beta_C$ is $\beta_C=0$, which leads us to the same upper bound on $k_{BC}$ as we obtained for the $\dot{z} > 0$ case, namely: 
\begin{equation}
z_c k_{BC}< \frac{a_A+(-a_B)}{2\sqrt{a_A}}~,
\end{equation}
but with the requirement $a_A<(-a_B)$.

Summarizing the results of this analysis:
\begin{itemize}
\item The kinematic upper bound is always
\begin{equation}
k_{BC}<\frac{a_A+(-a_B)}{2z_c\sqrt{a_A}}~,
\label{eq-boundk}
\end{equation}
which takes exactly the same form as in Eq.~(\ref{eq-kbound})
\footnote{Radiation leaks have been ignored in the AdS analysis but it is easy 
to see that restoring them reproduces Eq.~(\ref{eq-boundk}).}. In Minkowski space, this upper bound does indeed become arbitrarily large.
\item Although matchings with different signs of $\dot{z}$ looks quite 
different in Fig.\ref{fig-AdS}, they are continuously related situations.  
For light domain walls, $k_{BC} z_c <\sqrt{-a_B}$, we always get $\dot{z}<0$.  For heavier ones, $k_{BC} z_c >\sqrt{-a_B}$, we get $\dot{z}<0$ when $a_A<(-a_B)$, and $\dot{z}>0$ when $a_A>(-a_B)$.
\end{itemize}

\subsection{Escaping the crunch}
\label{sec-escapeAdS}  

Now that we understand how the matching works, we turn our attention to two interesting questions.
\begin{itemize}
\item Can we escape from the AdS crunch in the colliding bubbles by creating a classical transition to a lasting region with $H_C^2>0$?
\item Can the AdS crunch play a role, much like the initial singularity was important for `creating a universe in the lab'~\cite{FarGut87}, possibly allowing for solutions with $H_C^2>H_A^2$?
\end{itemize}
Recall that the way to guarantee a lasting region of some vacuum $C$ is to create heavy domain walls, and hence Repulsive geometries. When $a_A<(-a_B)$, the heavy walls will have $\dot{z}<0$. In this case, the region of $C$ will either end in a cosmological crunch, or result in an Oscillatory geometry {\em regardless of the value of the tension} $k_{BC}$. We prove this claim in Appendix~\ref{app-turnaround}. In this way, the large-scale structure of spacetime steps in to enforce limits on the types of classical transition geometries that can be produced -- Repulsive geometries simply don't exist, no matter what the tension is.

The story is potentially different when $a_A>(-a_B)$, where the heavy walls will have $\dot{z}>0$. In this case the necessary condition to form a lasting region of $C$ becomes the same as in the de~Sitter or Minkowski collisions---we need a Repulsive or Marginally Repulsive geometry. Again we prove that the Repulsive case is kinematically allowed analytically, and leave the Marginally Repulsive case to numerics. One key ingredient here is a large tension, $k_{BC}$, which indeed can become unbounded if $H_A^2=0$ as shown in Eq.~(\ref{eq-boundk}). On the other hand, the AdS region $B$ also increases the threshold for the Repulsive geometry,
\begin{equation}
k_{BC}^2>H_C^2 + |H_B|^2~.
\end{equation}
Combining this with Eq.~(\ref{eq-boundk}), we get
\begin{equation}
H_C^2<\frac{z_c^2(H_A^2-|H_B|^2)^2-4|H_B|^2}{4(1+z_c^2H_A^2)}~.
\end{equation}
It is easy to see that $H_C^2>0$ is quite achievable, actually quite natural 
for large $z_c$. So our answer to the first question is {\bf yes!}  A 
classical transition provides a way to escape from the AdS crunch, at least
for some observers.  

For the second question, we further demand that region $C$ has higher energy density than the false vacuum $A$ phase:
\begin{equation}
H_A^2<H_C^2<\frac{z_c^2(H_A^2-|H_B|^2)^2-4|H_B|^2}{4(1+z_c^2H_A^2)}~.
\end{equation}
A few lines of algebra leads us to
\begin{equation}
(-a_B)>3a_A~. \label{eq-strong}
\end{equation}
Note that $(-a_B)>a_A$ is enough to guarantee a matching with $\dot{z}<0$, 
which cannot produce a lasting region.  Eq.~(\ref{eq-strong}) exceeds that
threshold by a factor of 3, which strongly suggests that a Marginally Repulsive
geometry will not work either, a statement supported by numerics. Therefore, the 
structure of the spacetime never allows a lasting region of $C$ to be produced when
$H_C^2 > H_A^2$.

\section{Discussion and conclusions}\label{sec:conclusions}

Classical transitions arising from the collision between vacuum bubbles provide an alternative way of populating regions of the landscape. In this paper, we have found that gravitational effects allow for classical transitions to populate a much wider variety of vacua than simple field dynamics suggests. In particular, it is possible to produce regions of vacuum with a higher energy density than that of the colliding bubbles. Even collisions between crunching bubbles can produce a lasting region of inflating vacuum via a classical transition. What could these findings mean for populating vacua in eternal inflation?

One important aspect of classical transitions is that vacua which cannot be accessed by bubble nucleation can be reached through a classical transition. For example, considering only bubble nucleation, an AdS minimum that separates two sets of inflating vacua (in a one-dimensional potential) can carve the potential landscape into two disconnected islands~\cite{Clifton:2007en}. Classical transitions can provide a bridge between such islands, allowing a wider variety of vacua to be populated during eternal inflation. 

Even for vacua that can be reached through bubble nucleation, classical transitions might be the dominant production mechanism. This could occur if the bubble nucleation rate directly to the vacuum of interest is quite slow, but the rate to an adjacent minimum is relatively fast. Collisions between the quickly-nucleating bubbles might produce transitions to the slowly-nucleating vacua much more efficiently than direct bubble nucleation. Estimating the probability for a classical transition to occur as the joint probability of two bubbles nucleating in the same Hubble patch within the same Hubble time (so that they collide), this occurs roughly when the nucleation rates per unit four volume satisfy $\Gamma_{AB}^2 > \Gamma_{AC}$. 

In both of these cases, classical transitions have the potential to play an important role in the measure question for eternal inflation. The starting point for most measure proposals is a matrix of transition rates between the vacua in a landscape, which is then used to determine a regulated distribution of volume, bubble number, or some other quantity that can ultimately be tied to the relative number of observers in different vacua (see e.g.~\cite{Aguirre:2007gy,Guth:2007ng} for a review). Altered transition rates notwithstanding, depending on the choice of measure, classical transitions can drastically affect the overall picture of eternal inflation. For example, if one chooses a particular bubble nucleated during eternal inflation, all but a set of measure zero of the volume on a constant FRW time slice inside the bubble will be to the future of a collision~\cite{Garriga:2006hw}. If collisions can produce a classical transition to a lasting region of some new vacuum, then effectively all of the volume on any constant time inside the bubble will be removed, and replaced by the new vacuum. This is true both in the case of terminal (AdS or stable Minkowski) and non-terminal (de Sitter and unstable Minkowski) vacua. We hope to explore the implications of these and other issues in future work.

Regarding the ``no universe in a lab'' conjecture\cite{FarGut87,Vachaspati:1998dy}, our geometries provide some new insights. It is not easy to find explicit geometries without spherical symmetry, which are easily calculable, and support the conjecture. The fact that classical transitions with $H_C > H_A$ cannot occur from the collision of two bubbles, a fairly non-local and non-spherically symmetric initial condition, should dramatically boost our confidence on the conjecture.

Another area for future work regards the Oscillatory geometry. In our analysis, we have found that energy conservation allows for many oscillations, and at late-times the region undergoing the classical transition acts as an effectively Repulsive domain wall\footnote{This has been explored in the braneworld scenerio with 1 higher dimension\cite{BlaBuc03}}. This could lead to interesting observational signatures of the collision, since there is a somewhat stable topological defect that can persist during inflation.

\begin{acknowledgments}
The authors wish to thank A. Aguirre, L. Hui and E. Lim for helpful conversations. We extend our gratitude to R. Easther and E. Lim for organizing the ``Bubble Collisions and Lattice Simulation" workshop at Columbia University where this work was initiated, and A. Brown for invaluable input at the early stages of this project. MJ's research is funded by the Gordon and Betty Moore Foundation, and thanks Columbia University for their hospitality while portions of this work were completed. ISY's research is supported in part by the US Department of Energy, and also thanks California Institute of Technology for their hospitality.
\end{acknowledgments}

\begin{appendix}

\section{Details of the Oscillatory solutions}
\label{app-oscillate}

In this appendix, we treat the properties of the Oscillatory solutions in more detail. As we discussed in Sec.~\ref{sec-dS}, the initial collision between bubbles of the $B$ phase can give rise to a classical transition to the $C$ phase. In the Oscillatory geometries, the post-collision domain walls enclosing the $C$ phase eventually re-collide. At the second collision, the field dynamics suggests that if a classical transition occurs, it will be to the $A$ phase. This region can expand and re-collapse, possibly giving rise to a classical transition back to the $C$ phase, and so on, producing a pattern of classical transitions $C \rightarrow A \rightarrow C \rightarrow A \ldots$. Imposing energy conservation at each collision, we can numerically evolve the equations of motion on each segment, and determine the full solution. The main conclusions we draw from this set of numerical simulations are
\begin{itemize}
\item If the walls separating B from C and B from A have different tensions, there will be a finite number of oscillations. Eventually, there will not be enough center of mass energy in a collision to produce the heavier set of walls. 
\item When the oscillations terminate, there are two possibilities. First, there could be continued oscillation between the same vacuum: $C-A-C-A-A-\ldots$ or $C-A-C-C-\ldots$ depending on which set of walls is heavier. Second, there could be energy loss through an ejected shell of radiation. In this case, eventually the two colliding bubbles merge.
\item If the walls can oscillate for long enough, the region of $C$ and the walls that enclose it become one effectively repulsive domain wall: that is, $\dot{x}_B < 0$, and the walls are always moving out of the B phase in both of the colliding bubbles. 
\end{itemize}

To begin, we index the collisions by $i = 1, 2, 3 ,\ldots$, where $i = 1$ is the initial collision between the bubbles of $B$, and $i \geq 2$ label the subsequent collisions of the Oscillatory geometry. We must impose energy conservation Eq.~(\ref{eq-energy}) at each collision, which for $i \geq 2$, and neglecting the presence of any other energy sinks, is 
given by
\begin{equation}
\frac{\sqrt{a_{C_{i-1}}a_{C_i}}}{a_B}=
\bigg(\frac{-\beta_{C_{i-1}}+\sqrt{\beta_{C_{i-1}}^2+a_{C_{i-1}}}}
{-\beta_{B_{i-1}}+\sqrt{\beta_{B_{i-1}}^2+a_{B_i}}}\bigg)
\bigg(\frac{-\beta_{C_i}+\sqrt{\beta_{C_i}^2+a_{C_i}}}
{-\beta_{B_i}+\sqrt{\beta_{B_i}^2+a_{B_i}}}\bigg)~.
\end{equation}
Each of the quantities are evaluated at the position of the $i$-th collision $z = z_{c_i}$ (by symmetry, all collisions will occur at $x_C = 0$). The metric coefficient $a_{C_i}$ generically denotes the region undergoing the classical transition due to the $i$-th collision, and is given by
\begin{equation}
a_{C_i} = 1 - \frac{2 M_i}{z} + H_C^2 z^2, \ \ ({\rm for} \ i \ {\rm odd}), \ \ \ \ a_{C_i} = 1 - \frac{2 M_i}{z} + H_A^2 z^2, \ \ ({\rm for} \ i \ {\rm even})~.
\end{equation}
To satisfy energy conservation, the mass parameter $M_i$ must be different for each classical transition. Conservation of energy at the $i$-th collision then provides an equation for $M_i$ in terms of $M_{i-1}$. The definitions of $\beta_{B_i}$ and $\beta_{C_i}$ are as in Eq.~(\ref{eq-betas}), with $a_C \rightarrow a_{C_i}$. 

To find the Oscillatory geometries, we solve for energy conservation, numerically evolve the equations of motion for the wall to find the location of the following collision, and iterate. The first observation is that the outgoing velocity of the wall, as measured by $\dot{x}_C$, decreases each time a collision occurs (hence $\beta_C$ also continually decreases). This implies that the collisions become more frequent with increasing $z$, and that each subsequent classical transition produces a region whose maximum size along the $x_C$ direction is continually decreasing. Recalling that $\beta_B = \beta_C + k^2 z^2$, as $\beta_C$ (which must be negative) decreases in magnitude, eventually $\beta_B$ must become positive. When this occurs, the walls move away from the interior of both the $B$ bubbles, just as for the Repulsive geometries described in Sec.~\ref{sec-null}. The region undergoing the classical transition acts as an effective domain wall between the two bubbles, which at late times has a Repulsive behavior. In addition, because the outgoing velocity is continually decreasing, the effective center of mass energy available in each collision is decreasing. Therefore, if the tensions $k_{AB}$ and $k_{BC}$ are different, then the total number of oscillations must be finite. Eventually, the center of mass energy will become so small that it will be impossible to produce the heavier set of walls. 

\begin{figure}
\includegraphics[width=8cm]{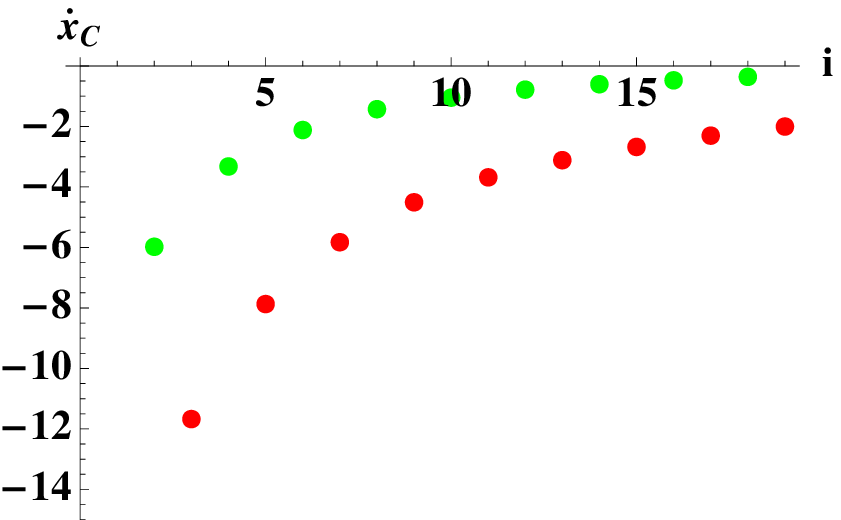}
\includegraphics[width=8cm]{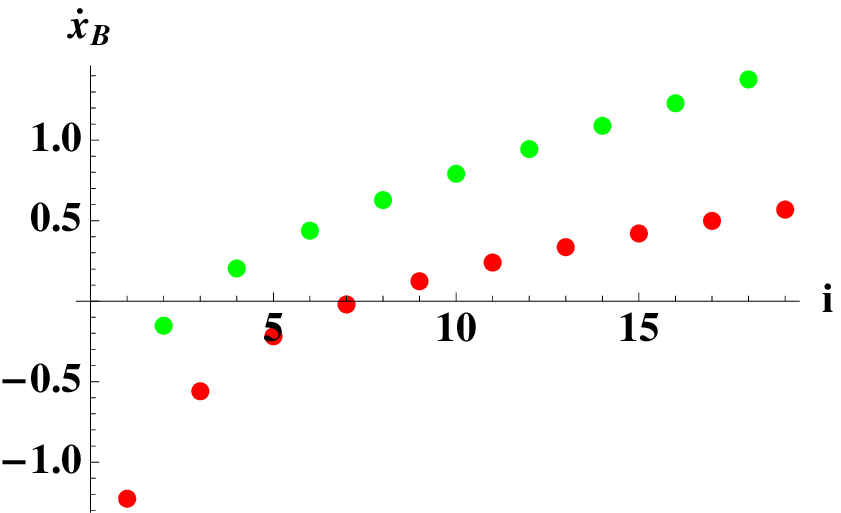} 
\caption{The outgoing velocity $\dot{x}_C$ (left) and $\dot{x}_B$ (right) for a number of collisions $i=1, 2,  \ldots 19$ in the Oscillatory geometry. This simulation was for the parameters (in units where $H_A = 1$) $\{R_0 = .33, \Delta x = 2.2, H_C = .4, H_B = .01, k_{AB} = .05, k_{AB} = .075 \}$. Note that the envelope of $\dot{x}_C$ decreases with $i$. The classical transitions must either terminate, or only involve vacuum $C$ for $i>19$. For $i>7$, $\dot{x}_{B}>0$, and the walls enclosing the region of $C$ move out the interior of both bubbles of $B$. 
\label{fig:xdotfigs}
}
\end{figure}

All of these behaviors are observed in numerical simulations. An example is shown in Fig.~\ref{fig:xdotfigs}, where we plot the outgoing value of $\dot{x}_B$ and $\dot{x}_C$ for a number of collisions.  Here, the tensions $k_{AB}$ and $k_{BC}$ are not equal, with $k_{BC} < k_{AB}$. The red points correspond to classical transitions to the $C$ phase (for the collision index $i$ odd), while the green dots are for classical transitions to the $A$ phase (for the collision index $i$ even). It can be seen that $\dot{x}_C$ oscillates, due to the fact that $k_{BC} < k_{AB}$, but the envelope continuously decreases in magnitude. For the collision $i = 18$, $\dot{x}_C$ is nearly zero, and there is no solution to the equations for energy conservation at $i=20$. The classical transitions must either terminate at this point, or transitions must occur to the $C$ phase only. Studying the plot for $\dot{x}_B$, it can be seen that for $i>7$, $\dot{x}_B$ is positive. Therefore, for the duration of the subsequent oscillations, the walls are continuously moving away from the interior of both bubbles of $B$.

Interestingly, in some potentials the Oscillatory solution must last forever.
The simplest example is a special case of the model presented in~\cite{HawMos82a}, a single scalar
field with a $Z_2$ symmetric potential that has a false vacuum $A$ in the middle, $B_1$ to the left and $B_2$ to the right.  The $Z_2$ symmetry implies that $H_A>H_{B1}=H_{B2}$ and $k_{AB1} = k_{AB2}$.  The classical transition in this special case will produce an oscillating region of the $A$ vacuum. Because $B_1$ and $B_2$ are different vacua, there must always be something separating the interiors of the two colliding bubbles. The classical transition can play this role if the oscillation does not end, and gives rise to a stable configuration in field space that goes between the position of vacuum $B1$, through $A$, to the position of vacuum $B2$. As above, such a configuration will move out of both colliding bubbles at late times, producing a Repulsive geometry.

\section{Collisions between different bubbles}
\label{app-B1B2}

In this appendix, we consider the case where the colliding bubbles that produce a classical transition are different. The equation of motion for the pre-collision and post-collision domain walls are identical to those presented in Sec.~\ref{sec-dS}. However, the equations of motion will be different for each of the four walls, and we must evolve them separately. The condition for energy conservation will also be different. We suppress the full expression, but it can be obtained as a straightforward generalization of the analysis in Refs.~\cite{Freivogel:2007fx,Aguirre:2007wm,Chang:2007eq}. In the limit of null walls, energy conservation becomes
\begin{equation}\label{eq:unequalnull}
a_A a_C = a_{B1} a_{B2}~, 
\end{equation}
where 
\begin{equation}
a_{B1} = 1+H_{B1}^2 z^2, \ \ \ \ a_{B2} = 1+ H_{B2}^2 z^2~,
\end{equation}
are the metric functions in each of the two colliding bubbles (where $B1$ denotes the bubble nucleated at $x<0$ and $B2$ the bubble nucleated at $x>0$), with $a_A$ and $a_C$ defined as before. Also as before, we can define the functions
\begin{equation}
\beta_{C1} = \frac{a_C - a_{B1} - z^2 k_{B1C}^2}{2 z k_{B1C}}, \ \ \ \beta_{C2} = \frac{a_{B2} - a_C + z^2 k_{B2C}^2}{2 z k_{B2C}}~,
\end{equation}
where the physical criteria for a classical transition to occur is
\begin{equation}
\beta_{C2}-\beta_{C1} > 0~.
\end{equation}
Substituting for $a_C$ using Eq.~(\ref{eq:unequalnull}), we have in the limit of large $z_c$ that
\begin{equation}
\frac{1}{2 k_{B2C}} \left( H_{B2}^2 + k_{B2C} -H_{B1}^2 \frac{H_{B2}^2}{H_A^2}  \right) > \frac{1}{2 k_{B1C}} \left( H_{B1}^2 \frac{H_{B2}^2}{H_A^2} - H_{B1}^2 - k_{B1C}^2 \right)~.
\end{equation}
For $H_A^2 > H_{B1,\  B2}^2$, which we require to have vacuum bubble solutions, the left hand side is positive definite, while the right hand side is negative definite. Therefore, in the null limit there can always be a classical transition. 

The classes of geometries that can be produced are similar to those produced in the collision between identical bubbles. 
\begin{itemize}
\item If both walls are accelerating away from $C$, (which includes being Repulsive or accelerating into $B_i$), then region $C$ is lasting. 
\item If $H_C^2-H_{Bi}^2 \gg k_{BiC}^2$, then the geometry will be Oscillatory. However, in this case, the subsequent collisions will not occur at $x=0$, but will initially drift towards the vacuum of higher energy. This can be understood by looking at the z-position of the turning point in $x$ for both walls:
\begin{equation}
z_{1}^3 = \frac{2 M}{H_C^2 - H_{B1}^2 -k_{B1C}^2}, \ \ \ \ z_2^3 =  \frac{2 M}{H_C^2 - H_{B2}^2 -k_{B2C}^2}~.
\end{equation}
To the extent that the tensions can be ignored, it can be seen that the turn-around on the side with a lower $H_{B1}^2$ will happen first. This means that the wall initially moves towards the bubble containing higher energy density. Eventually, just as for identical bubbles, we will have $\dot{x}_B > 0$, and the wall will accelerate out of both bubbles. Also, just as for identical bubbles, since the tensions on either side of the region that has undergone the classical transition are not equal, the oscillations must terminate.
\item When the two walls behave differently, the general fate of region $C$ 
depends on the integral
\begin{equation}
\int \frac{\beta_{C2}-\beta_{C1}}{a_C}~dz~.
\end{equation}
If it ever goes to zero, the two walls recollide.  If not, it is a lasting 
region.  Nevertheless there will still be a general drift into the region with
higher $H_{Bi}^2$.
\end{itemize}

\section{General geometries with $\dot{z}<0$ matchings}
\label{app-turnaround}

\begin{figure*}
\begin{center}
\includegraphics[width=8cm, angle=0]{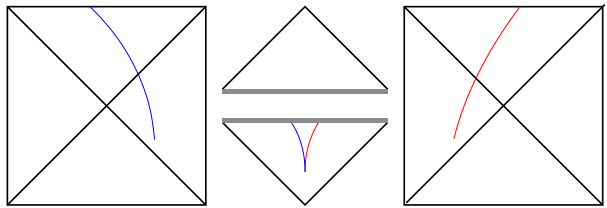}
\end{center}
\caption{The two triangles in the middle are the full coordinate coverage for 
Hyperbolic-Schwarzschild with $M<0$. The $\dot{z}<0$ matching to AdS bubbles
$B$ on both sides has to be in the lower triangle, which ends up in a crunch
in the same time as the AdS crunch outside.}
\label{fig-crunch}
\end{figure*}

In this appendix we will show that $\dot{z}<0$ matchings can never have a 
lasting region $C$. As shown in Sec.\ref{sec-null}, the null limit of the energy conservation 
equation is a useful tool to find qualitatively different geometries.  
\begin{equation}
1-\frac{2M_C}{z_c}+H_C^2z_c^2=\frac{(1-|H_B|^2z_c^2)^2}{1+H_A^2z_c^2}~.
\label{eq-Mc}
\end{equation}
The parameter $M_C$ can be either positive or negative, and be completely consistent with energy conservation. In fact, for $\dot{z} < 0$, since we require $\beta_C > 0$, the mass parameter {\em must} be negative for heavy domain walls. 
Geometries with $M_C>0$ are similar to the Hyperbolic-Schwarzschild in 
Fig.~\ref{fig-AdS}, while the $M_C<0$ ones are usually drawn as an 
up-side-down triangle with a future singularity.  In fact it is equally legal
to have a triangle with a past singularity, as in Fig.~\ref{fig-crunch}, and a 
matching with $\dot{z}<0$ corresponds to a portion in this region.  In these kinds 
of matchings, one always ends up with a cosmological crunch in the region of $C$, 
similar to what occurs inside of AdS bubbles. 

As for $M_C>0$, the geometries fall into two classes. Let $Z_C$ be the horizon
in the $C$ region (note, this is {\em not} the cosmological event horizon, but the $z$-position where $a_C = 0$), 
Eq.~(\ref{eq-Mc}) shows that both $Z_C>|H_B|^{-1}$ and 
$Z_C<|H_B|^{-1}$ are possible. They correspond to slightly different 
geometries.

When $Z_C<|H_B|^{-1}$, the domain wall crosses the horizon in region $B$ 
first. After that $a_C$ is positive and decreasing, $a_B$ is positive and 
increasing, eventually we reach a $\tilde{z}>Z_C$ where
\begin{equation}
a_C(\tilde{z})=a_B(\tilde{z})+\tilde{z}^2k_{BC}^2~.
\end{equation}
By Eq.~(\ref{eq-betas}), this means $\beta_C$ has to change sign here. The domain walls enclosing the region of $C$ turn around before crossing the horizon in region $C$, and are doomed to re-collide, forming an Oscillatory geometry.

When $Z_C>|H_B|^{-1}$, the domain wall first crosses the $C$ horizon to where
both $a_C<0$ and $a_B<0$.  Now instead of $\beta_C$, Eq.~(\ref{eq:effpot}) 
shows that $\dot{z}$ turns around at $\tilde{z}>|H_B|^{-1}$ for which
\begin{equation}
-a_B(\tilde{z})=(\sqrt{-a_C(\tilde{z})}+\tilde{z}k_{BC})^2~.
\end{equation}
Clearly this is guaranteed to happen before crossing the $B$ horizon.  
Eq.~(\ref{eq:effpot}) also implies that the trajectory is time 
reversible.  So after $\dot{z}$ turns around, it will later enter the top 
quadrant of the Hyperbolic-Schwarzschild geometry and collide with the other
wall just like how they emerged initially. This is also an Oscillatory 
geometry.

\end{appendix}

\bibliography{gravclasstrans}

\end{document}